\documentclass[10pt,sigconf,letterpaper]{acmart}

\usepackage{enumitem}
\usepackage{tikz}
\usepackage{caption}
\usepackage{subcaption}

\usepackage{afterpage}
\usepackage[export]{adjustbox}
\usepackage{blindtext}
\usepackage{subcaption}
\usepackage{graphicx}
\usepackage{lipsum}
\usepackage{xspace}
\usepackage{soul,color}
\usepackage{booktabs}
\usepackage{multirow}
\usepackage{array,boldline}
\usepackage{makecell}
\usepackage{algorithm}
\usepackage{algpseudocode}
\usepackage{float}
\usepackage{mathtools}
\usepackage{nicematrix}
\usepackage{url}
\hypersetup{
    colorlinks=true,
    linkcolor=blue,
    filecolor=magenta,      
    urlcolor=cyan,
}
\usepackage[framemethod=TikZ]{mdframed}

\newcommand{\ballnumber}[1]{\tikz[baseline=(myanchor.base)] \node[circle,fill=.,inner sep=1pt] (myanchor){\color{-.}\bfseries\footnotesize #1};}

\AtBeginDocument{%
  \providecommand\BibTeX{{%
    \normalfont B\kern-0.5em{\scshape i\kern-0.25em b}\kern-0.8em\TeX}}}

\setcopyright{acmcopyright}
\copyrightyear{2018}
\acmYear{2018}
\acmDOI{10.1145/1122445.1122456}

\acmConference[Woodstock '18]{Woodstock '18: ACM Symposium on Neural
  Gaze Detection}{June 03--05, 2018}{Woodstock, NY}
\acmBooktitle{Woodstock '18: ACM Symposium on Neural Gaze Detection,
  June 03--05, 2018, Woodstock, NY}
\acmPrice{15.00}
\acmISBN{978-1-4503-XXXX-X/18/06}

\begin{document}


\title{PhishClone: Measuring the Efficacy of Cloning Evasion Attacks}


\author{Arthur Wong}
\email{z5205512@unsw.edu.au}

\affiliation{%
  \institution{UNSW \\ Cyber Security CRC}
  \country{Australia}
}

\author{Sharif Abuadbba}
\email{sharif.abuadbba@data61.csiro.au}
\orcid{1234-5678-9012}
\affiliation{%
  \institution{CSIRO's Data61 \\ Cyber Security CRC}
  \country{Australia}
}

\author{Mahathir Almashor}
\email{Mahathir.Almashor@data61.csiro.au}
\affiliation{%
  \institution{CSIRO's Data61\\ Cyber Security CRC}
  \country{Australia}
}

\author{Salil Kanhere}
\email{salil.kanhere@unsw.edu.au}
\orcid{1234-5678-9012}
\affiliation{%
  \institution{UNSW}
  \country{Australia}
}

\renewcommand{\shortauthors}{Trovato and Tobin, et al.}
\newcommand{\sharif}[1]{\textcolor{blue}{[#1 ---Sharif]}}
\newcommand{\arthur}[1]{\textcolor{purple}{[#1 ---Arthur]}}
\setcopyright{none}
\settopmatter{printacmref=false} 
\renewcommand\footnotetextcopyrightpermission[1]{} 
\pagestyle{plain}

\begin{abstract}

Web-based phishing accounts for over 90\% of data breaches, and most web-browsers and security vendors rely on machine-learning (ML) models as mitigation. Despite this, links posted regularly on anti-phishing aggregators such as PhishTank and VirusTotal are shown to easily bypass existing detectors.
Prior art suggests that automated website cloning, with light mutations, is gaining traction with attackers. This has limited exposure in current literature and leads to sub-optimal ML-based counter-measures. The work herein conducts the first empirical study that compiles and evaluates a variety of state-of-the-art cloning techniques in wide circulation.
We collected 13,394 samples and found 8,566 confirmed phishing pages targeting 4 popular websites using 7 distinct cloning mechanisms. These samples were replicated with malicious code removed within a controlled platform fortified with precautions that prevent accidental access. We then reported our sites to VirusTotal and other platforms, with regular polling of results for 7 days, to ascertain the efficacy of each cloning technique. Results show that no security vendor detected our clones, proving the urgent need for more effective detectors. Finally, we posit 4 recommendations to aid web developers and ML-based defences to alleviate the risks of cloning attacks.

\end{abstract}


\maketitle

\section{Introduction}\label{sec:intro} 

Phishing gains a user's trust by impersonating legitimate persons and organisations. This is then exploited to obtain sensitive credentials and personal or commercial data, which allows system compromise and on-selling to underground information markets. Popular and effective forms of phishing include spear-phishing in emails \cite{ho2017detecting,kashapov2022email,kim2020deepcapture,evans2021raider}, and web phishing \cite{shaikh2016literature}. Web phishing remains a prevalent and effective cyber threat, affecting millions \cite{verizon2018data} of users and costing nearly \$500 million every year \cite{mathews2018phishing}. This threat has escalated with the COVID pandemic forcing businesses to increase remote operations. This is made plain when we observe the rate of phishing attacks rising up to 220\% in the early months of 2020 \cite{bitaab2020scam}.

In response, the cyber-security community has developed various anti-phishing tools to detect web-phishing sites. The most common technique is that of user-reported URL block-lists, which are included in most major web-browsers \cite{sandhu2015google}. However, these lists can often lag behind the magnitude of new phishing sites being created \cite{peng2019opening,moore2008evaluating}. Recently, ML-based techniques have arisen to dynamically classify elements of a web-page as phishing or benign \cite{ubing2019phishing,corona2017deltaphish,xiang2011cantina+,shmalko2022profiler,abuadbba2022towards,almashor2021characterizing}.

In an ever deepening arms race, attackers have adopted more sophisticated mechanisms. Prior work has uncovered adversarial evasion attacks specifically created to avoid detection by ML-based classifiers \cite{lei2020advanced,corona2017deltaphish,laskov2014practical}. One concerning development is ``\textit{cloning}'', which has limited exposure in the current art. Here, an attacker copies the structure and characteristics of trusted and famed sites, which nullifies the ability of ML classifiers to accurately label such sites as malicious.

\noindent Thus, we present these research questions and contributions:

%

\begin{itemize}[leftmargin=*]
    \item \textbf{How prevalent is website cloning in phishing?} In section \ref{sec:identify}, our analysis found 14 clusters comprising of 49 sites that targeted Facebook, Microsoft, PayPal, and eBay. Each brand had at least 2 clusters, which reveals the versatility of cloning across multiple targets. Given their popularity, this represents an alarming risk for millions of users.
    
    \item \textbf{How do attackers utilize website cloning?} We further our investigation in Section \ref{sec:clusters}, where we exposed 7 distinct cloning methods used against current ML-based defenses. The attacks are deconstructed in a secure environment, and we provide practical insights into their use and limitations.
    
    \item \textbf{What is their effectiveness against commercial products?} We detail how our own cloning attacks were deployed within controlled environments in Section \ref{sec:deploy}, and present a discussion as to why none of the security vendors used by VirusTotal could detect our phishing sites.
    

    
\end{itemize}

Our aim is to aid the community in improving detection criteria, and assist developers to harden their sites against such cloning attacks. Thus in section \ref{sec:rec}, we conclude the work with practical recommendations for both web-developers and current detectors that may alleviate this threat potential.
\begin{figure*}[h]
  \centering
  \includegraphics[width=1.0\linewidth]{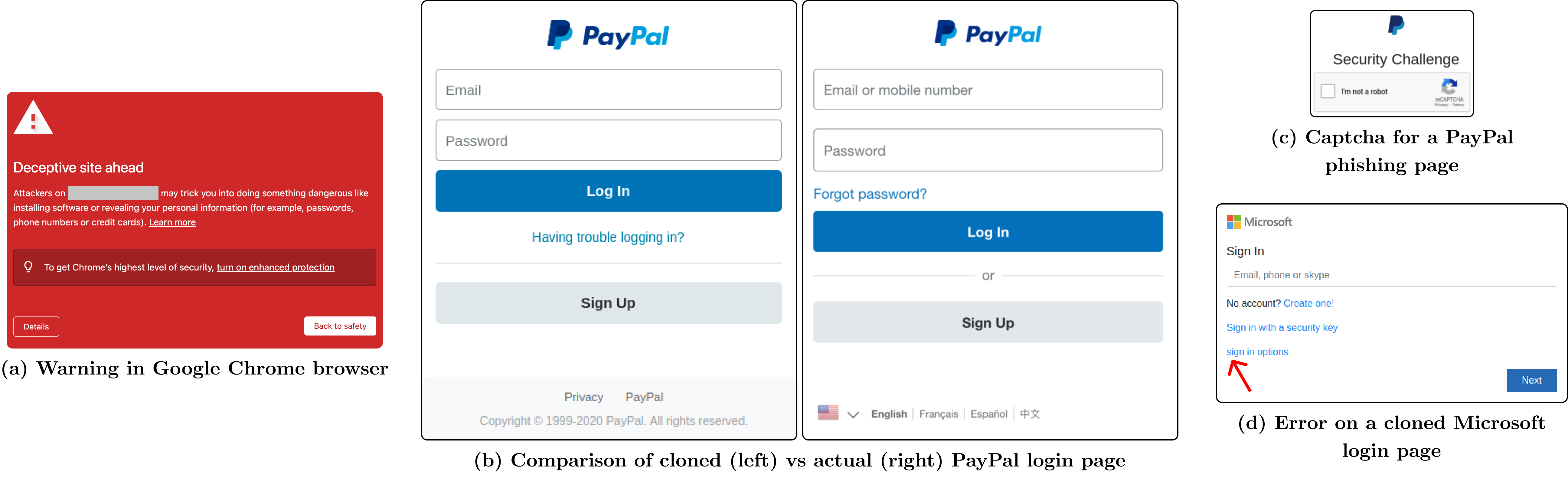}
  \caption{Gallery of screenshots seen by end-users when navigating to both phishing and legitimate sites.}
\label{fig:gallery}
\end{figure*}

\section{Background and Related Work}
This section provides prior knowledge for anti-phishing block-lists, defences, and website cloning.
\subsection{Anti-phishing Block-lists}
Anti-phishing block-lists are a common way of preventing
users from accessing phishing websites, and are integrated into most modern web-browsers. Figure \ref{fig:gallery}(a) shows an example of blocked access to a phishing website. Block-lists are constantly updated with URLs identified as malicious, with some like PhishTank \cite{bell2020analysis} using manual peer-review of suspect sites in order to classify them. VirusTotal \cite{peng2019opening} uses specialised security vendors, and the widely used Google Safe Browsing (GSB) \cite{sandhu2015google}, gathers defence system data \cite{whittaker2010large}.

\subsection{Machine-learning (ML) Defences}
With the scale of malicious websites on the Internet today, an automated solution is needed to protect users. Anti-phishing models analyse features like URL strings, and webpage Document Object Model (DOM), containing the structure of HTML elements in the page  and identify suspicious patterns \cite{li2019stacking}. 
Each element is weighted, depending on how likely it belongs to a phishing page. By aggregating the weights, the algorithm will output a decision score on how likely it is that the page is phishing. 
For example, Xiang et al. \cite{xiang2011cantina+} developed CANTINA+ to identify sites with the same features as previously detected phishing sites. As more phishing sites are identified, the software will adjust how it weights particular DOM elements~\cite{lei2020advanced,abu2007comparison}. 
Visual similarity can also be used to identify potential attacks. Many works computed the similarity between suspicious and legitimate sites, based on the difference in size, positioning for blocks of elements, or logos ~\cite{zhang2013web,ebrahimzadeh2014efficient,bozkir2020logosense,lin2021phishpedia}. Unfortunately, visual-similarity methods lack support for lesser-known pages. Attackers who target relatively unknown websites can overcome visual similarity-based defenses that compare sites to well-known brands. Another option is to analyse the URL of a phishing site. Hung et al. \cite{le2018urlnet} applied deep learning to capture semantic information hidden in URLs. However, attackers can bypass this analysis using random domains or shortening their URLs.
\subsection{Website Cloning}
Website cloning is the idea that an attacker can imitate a known website by using the HTML and/or imitating the appearance of that site. One form of cloning is where sites will take resources from their target (like images, stylesheets, and HTML structure) and use those to try and appear genuine. Figure \ref{fig:gallery}(b) shows an example of a blocked cloned site, that has taken the visual appearance of a legitimate site. Some sites will attempt to copy the visual appearance of a target site, while others will be more aggressive and use similar domains, similar assets, or make similar network requests. Because anti-phishing classifiers look through the DOM structure for suspicious elements, cloning a friendly website will theoretically provide the malicious site with the same classification score as their target, especially for less-known websites that are not protected by the visual similarity defences. 

Many studies investigated how phishing pages avoid detection. Oest et al. \cite{zhang2021crawlphish} analysed 100K phishing websites, and found that 31.31\% of them implemented client-side cloaking mechanisms. Nevertheless, cloning attacks in particular have limited exposure in current literature. Hence, our research is the first empirical study exploring the prevalence and methodology of website cloning. 
\section{Identifying cloning techniques}\label{sec:identify}
How prevalent is the use of website cloning? To answer this, we firstly collected and examined phishing websites and reverse-engineered their source code as necessary. We then used clustering and visual DOM comparison to link them to their legitimate websites as explained below.

\subsection{Dataset Composition}
PhishTank was chosen as a suitable repository of phishing sites, as they have over 12,000 URLs at any given time, reported by users \cite{bell2020analysis}. We scraped a collection of 13,394 URLs and HTML files from February to April 2021. ML defences may not have analysed these URLs, as they are added to PhishTank based on manual user reports. This requires us to submit them to security vendors to gather data on how automatic anti-phishing systems classify them.

Cloned attacks choose a well-known target brand to imitate to maximise their chances of gaining the users recognition and trust \cite{geng2015combating}. We focused our study on four brands: \textit{Facebook, Microsoft, PayPal} and \textit{eBay}. To restrict our data to these brands, we filtered the dataset by searching for the brand names in the HTML source code.  In total, we analysed 8,566 titles across our four targets. Table \ref{tab:target-samples}shows how many samples we collected for each of our targets after the filtering.

\begin{table}[h]
\begin{center}

\begin{tabular}{r|l|l}

     \textbf{Target/Victim Brand} & \textbf{Sample Count} & \textbf{Cloned sites}\\
     PayPal & 3330 & 20 \\
     Microsoft & 1960 & 11\\
     Facebook & 2847 & 6\\
     eBay & 429 & 12 \\
     Other & 4828 & - \\ 
     Total & 13,394 & 49 \\

\end{tabular}

\end{center}
\caption{PhishTank samples for targeted brands.}
\label{tab:target-samples}

\end{table}

\begin{figure}[!h]
  \centering
  \includegraphics[width=1.0\linewidth]{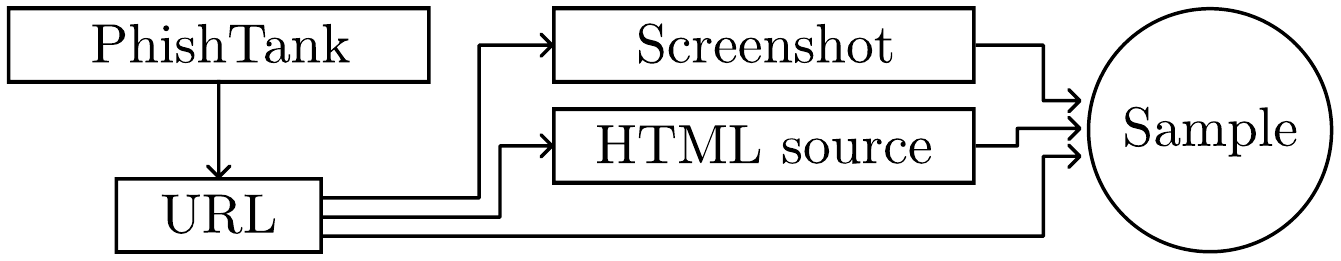}
  \caption{Artefacts were obtained for each sample.}
\label{phishing_sample}
\end{figure}

For each page, the visual appearance was collected to analyse the appearance of users. We used the Selenium package in Python to bypass Google Chrome security features and take a screenshot on a browser. Figure \ref{phishing_sample} illustrates our data collection for each phishing URL. 

 \noindent
 
\subsection{HTML Clustering}
To systematically study the samples, we ran a clustering algorithm over all of the HTML files for each brand, grouping together websites using similar cloning techniques. HTML documents can have many elements repeated, and the order of each element is important. This is why the SequenceMatcher \cite{rao2018characteristic} class was appropriate, since it allows us to identify contiguous subsequences of elements in a sequence. The DOM structure of each pair of pages was treated as a sequence of individual HTML elements to be compared. The attributes of each tag were ignored. The output of this algorithm was an image file containing screenshots of samples, grouped together with other similar samples.

The SequenceMatcher class runs a linear comparison for each pair of ordered sequences to find the total number of matching blocks. Each block is a sub-sequence of individual elements that are identical in both sequences. Using the number of matching blocks $m$ and the sum of the lengths of both sequences $t$, we computed the similarity $p$ of two sequences by calculating $p = 2m / t$, where $0 \leq p \leq 1$. This algorithm ran across all possible pairs of samples targeting one particular brand. Because web cloning involves copying as much of the benign webpage as possible, we expected that similarity scores would be high for pages that utilised cloning. After experimenting with our samples, we chose 80\% similarity as a threshold  score. 

\noindent{\bf{Findings.}} From our analysis, we found 49 websites, split into 14 clusters and 7 unique cloning mechanisms. Although only 0.5\% of websites targeting our 4 chosen brands used cloning, it should be noted that all four of our targets had at least 2 clusters, meaning this kind of attack is versatile across different targets. These brands have billions of users every month, meaning that small percentages of users being tricked by cloned sites could result in millions of accounts being compromised. Table \ref{table of clusters} shows how many clusters each cloning mechanism had targeting each brand. 
\begin{table}[!h]


\begin{tabular}
{
    r |
    c
    c
    c
    c
}

\textbf{Technique} &
\textbf{EB} &
\textbf{FB} &
\textbf{PP} &
\textbf{MS} \\[0.4em]

No-Code Solution              & - & 2 & - & - \\
Dynamic JavaScript Generation & 2 & - & 1 & 1 \\
{CAPTCHA-Blocking}            & - & - & 1 & - \\
Direct Copy                   & 1 & - & 1 & - \\
Hardcoded Data                & - & - & 2 & - \\
Typography Mistakes           & - & - & - & 2 \\
Fully-custom cloning          & - & - & 1 & - \\

\end{tabular}

\caption{Number of clusters found for each cloning mechanism targeting eBay (EB), FaceBook (FB), PayPal (PP), and Microsoft (MS).}

\label{table of clusters}

\end{table}

\section{Analysis of Cloning Techniques}\label{sec:clusters}
From the 49 phishing samples we found utilising cloning, we identify 7 different techniques used below, in order of popularity. We believe attackers have three aims with cloning: to easily produce an attack, mimic the appearance of the target site, and evade anti-phishing defenses. Each cloning mechanism aims to strike a balance between these objectives. \\

\begin{figure}[!h]
  \centering
  \includegraphics[width=1.0\linewidth]{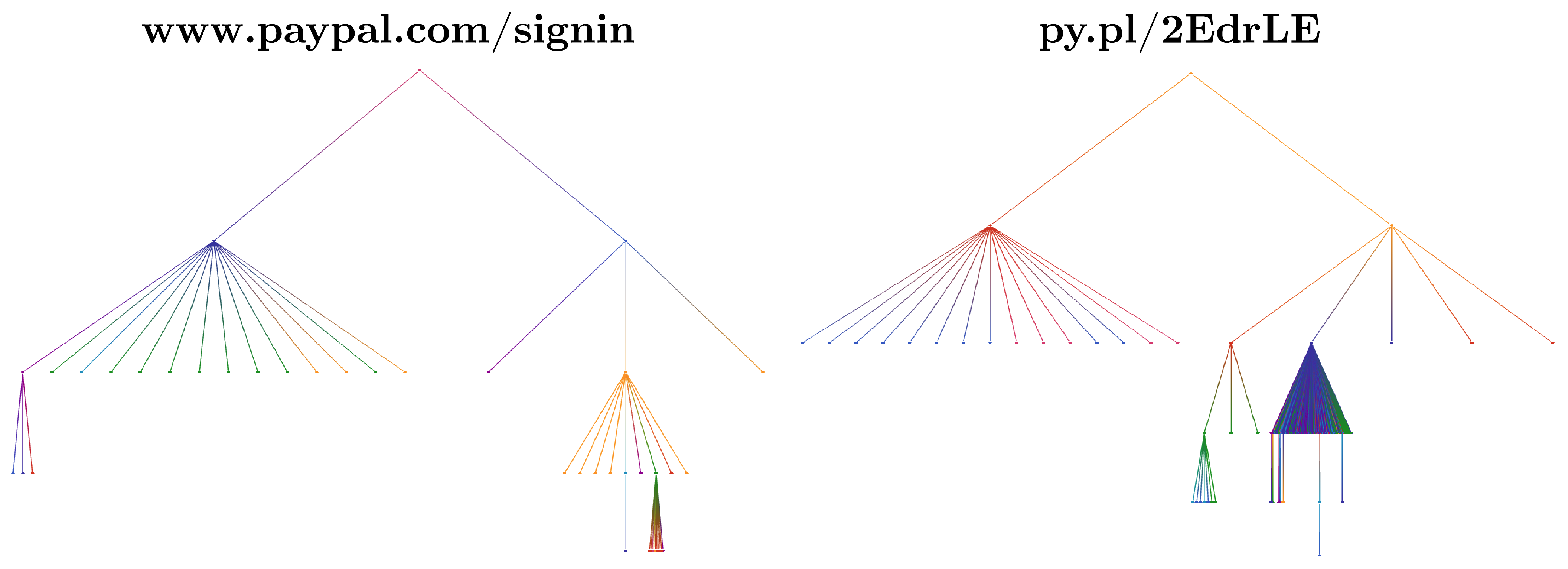}
  \caption{DOM structure comparison for Direct Copy.}
\label{fig:dom-tree-comparison}
\end{figure}

\noindent{\ballnumber{1} \bf{Dynamic JavaScript Generation.}} Some sites clone their target by generating the website with JavaScript, rather than by copying the HTML. The lack of HTML means ML-based defenses cannot properly classify the site and identify it as malicious. Some examples used the DOM API \cite{wood1998document} to interface with the elements in the browser. Other attackers chose to use web frameworks, such as ReactJS and Knockout.js. These frameworks provide more robust ways to implement complex web design, which may have the most success against anti-phishing defenses. This may explain why it is the most popular in our data, since our data was sourced from lists of user-reported sites containg sites that possibly already evaded automatic detection. \\
\noindent{\ballnumber{2} \bf{Direct Copy.}}  The simplest way to clone the site is to directly copy the target site. Attackers can obtain the source code for a benign website and copy the HTML, and host it on their own servers. Features on the benign site that are unusable on the copy, like resetting passwords or integrating with Google accounts, are usually removed. Our DOM analysis discovered that cloned sites have nearly identical <head> sections to the real site, which includes metadata and assets that don't directly add visual elements to the page. The limitations in copying are usually in deeper branches in the <body> section, which contain the features that cannot be cloned. Figure \ref{fig:dom-tree-comparison} compares the DOM structure of the real PayPal site, with a phishing site that directly copied the HTML. The root node represents the base document, and nodes to the left indicate elements that come first. This cloning mechanism saves time for the attackers, and assists in preserving the visual appearance of the site, which is necessary to deceive a user.\\

\noindent{\ballnumber{3} \bf{Hard-Coding Dynamic Data.}}
Modern web developers do not encode data directly in their HTML source code. Instead, data is fetched from an API call as part of the site execution.\cite{sohan2015case} This is typically done for data that can change over time, as modifying a database is much cheaper than rewriting and re-deploying HTML code to reflect updated information. It also improves page-load speeds, and makes the source code cleaner for developers. On the real PayPal site, the telephone codes for each country are stored on PayPal servers, and rendered at run-time when the page loads. There are 2 issues for the attackers here. (1) If the phishing pages were to include the API calls from the original PayPal website, they would contact PayPal servers and potentially alert their developers to malicious activity. (2) There are also security features like the Same-Origin policy \cite{schwenk2017same}, that prevent websites from interacting with data hosted on other servers. Hard-coding dynamic data within the phishing webpage itself helps attackers avoid these challenges.\\  
\noindent{\ballnumber{4} \bf{Typography Mistakes.}}
Since text forms part of an HTML element, altering text elements slightly will change the keywords that appear in an ML algorithm. The site still retains a similar visual appearance, and may still be picked up by ML algorithm that looks at screenshots of websites to evaluate them \cite{lin2021phishpedia}. Many cloned sites included small mistakes in spelling or grammar. Figure \ref{fig:gallery}(d) shows an example with the words ``sign in options'' missing a capital letter at the start of the sentence. This would work well against users whose first language isn't English, and may not spot the errors. ML algorithms might search for popular keywords like ``login'', ``password'' and ``credit card'' to determine if a site is asking for sensitive user information \cite{l2010latent}\cite{xiang2011cantina+}\cite{ding2019keyword}. However, changing the spelling of a word can help the attack avoid detection, while still appearing legitimate to real users.\\ 
\noindent{\ballnumber{5} \bf{CAPTCHA-blocked site.}}  Some attacks combined cloning and cloaking mechanisms. Maroofi et al. \cite{maroofi2020you}, explored how websites with CAPTCHA could be used to stop automatic anti-phishing software analysing a page. However, we found that attackers take this one step further by directing the user to the original CAPTCHA that  exists on the target website, as seen in Figure \ref{fig:gallery}(c), then redirect them into their phishing site. The resulting cloned phishing page closely resembles the real PayPal site that is familiar to users, while simultaneously evading anti-phishing systems.\\
\noindent{\ballnumber{6} \bf{Fully-custom Copy.}}
Some samples retained an identical appearance to their target; Nevertheless, a visual analysis of the DOM tree showed that none of the HTML elements had been copied over. The sites still used assets from their target pages, such as CSS files and images. Having completely different HTML would circumvent ML algorithms, as there is no resemblance to the target site, or any other phishing site. This type of attack would take a significant amount of time to produce, but would also be difficult to detect.\\
\noindent{\ballnumber{7} \bf{Using no-code Solutions.}} There are many online services that allow anyone to create websites using drag-and-drop templates, with famous examples being WIX  and SquareSpace. Attackers can use these templates to create a page that tricks users unfamiliar with the real service. This cloning mechanism is extremely easy to implement, and the service used to create the site often provides hosting, eliminating the cost of running a server. 

\section{Deploying cloned sites}\label{sec:deploy}
We aim to answer the question: \textit{how effective cloning mechanisms are against current machine-learning models?} Our experiments generated and deployed sites using the 7 different cloning mechanisms. We obtained ethical clearance from Paypal to target their login page and ran our tests under a controlled environment. We underwent the following steps: (1) Generating samples to replicate cloning attacks; (2) Hosting our samples in a safe manner that is similar to how attackers host phishing pages; (3) Continuously submitting our websites to security vendors, ensuring they are scanned by anti-phishing defenses; (4) Evaluating the effectiveness of each cloning mechanism by observing how long they remain off of blacklists.

\noindent{ \bf{Ethical Consideration.}} We received ethical clearance from PayPal to use their login page as a target for our replica cloning attacks. Our implementation had to meet certain conditions, to ensure the security and privacy of any real users were not jeopardised. (1) Our sites could only be targeted at anti-phishing APIs, and not users. (2) We used secure HTTPS. (3) The URLs had to be randomised. (4) The URLs of the sites could not be circulated by email. (5) Our sites had to be dummy login sites with no data collection. These conditions eliminated the risk of any real users mistaking our site as legitimate.\\
\noindent{\textbf{Generating Samples.}} We used the source code for the PayPal login page, including the HTML and CSS files, as a base for each of the 7 cloning techniques. Our hosted sites were made to be as similar as possible to the attacks that we observed, while following the conditions of the ethical agreement.\\
\noindent{\textbf{Deploying.}} We used the Netlify service to deploy 6 of our 7 sites. Netlify created randomised URLs for us, satisfying the condition of our ethical agreement. To maximise the chances of our sites being properly scanned by VirusTotal vendors, each sample was hosted under 4 different domains simultaneously. The last site was a no-code solution implemented with Wix, the same service that many of our phishing samples were using. WIX provided a free hosting service and a domain name of the structure "xxxxx.wixsite.com/my-site", where xxxx was our random account username. A study into the hostnames of sites on VirusTotal found that 46.5\% of malicious sites were hosted on domains that were provided by an external service (e.g. "webhost000.com") \cite{de2021compromised}. We were able to reflect this by hosting our sites on an external service with the name "Netlify" in the URL.\\
\noindent{\textbf{Scanning VirusTotal.}} Once the sites were hosted, we used 4 virtual machines for polling the VirusTotal scanning API. Each machine submitted a report for every sample under a different URL. Using separate machines increases the chances that VirusTotal vendors will analyse our pages and return results. Our script would send a request to each of the public APIs every half an hour for each URL, with a 20 second pause between each individual URL request to ensure we did not go over the request limit for our API key. VirusTotal provides an aggregate rating, based on the results of their security vendors. Because the vendors need time to contact VirusTotal and provide their results, there is a delay between submitting a link to VirusTotal for scanning, and the results updating. We also polled GSB and Sucuri separately, as they were identified as potentially having different labels when polled individually to those provided to VirusTotal \cite{peng2019opening}. As Sucuri did not provide an API, we submitted our URLs to their web interface twice daily. 
 
Previous studies into VirusTotal have shown that it takes 4 days for URL phishing labels to update \cite{peng2019opening}. A study into the file-scanning API showed that the minimum waiting period was 5 days, with the 5 most reputable vendors having a stable label within 1 day for 88.05\% of files submitted \cite{zhu2020measuring}. From this, we decided that waiting 5-7 days was a reasonable amount of time for the label to settle. For our experiments, we deemed a site as ``detected'' if its URL was flagged by any VirusTotal (VT) vendors, or either of the Sucuri (SU) or Google Safe Browsing (GSB) engines. 

\section{Effectiveness of cloning}\label{sec:effecacy}
After our sites were hosted for 7 days, we studied their detection rates and suggested possible reasons for our results.\\

\noindent{ \bf{Results Summary.}}
Table \ref{tab:results} shows the detection rates. After waiting 7 days, we found no detection for any cloning mechanism, from any security vendor. This result indicates that current anti-phishing models  are unable to properly classify sites using our cloning mechanisms. This means that phishing sites using these mechanisms could be live and evade being blacklisted, putting millions of users at risk.


\begin{table}[!h]
\centering
  
 \begin{tabular}{%
    r|%
    l%
 }
    \textbf{Cloning Technique} &
    \textbf{VT, SU and GSB} \\

    Dynamic JavaScript Generation &
    Not detected \\

    CAPTCHA-blocked Site &
    Not detected \\
    
    Direct Copy &
    Not detected \\

    Hard-coding Dynamic Data &
    Not detected \\

    Typography Mistakes &
    Not detected \\

    Fully-custom Copy &
    Not detected \\

    No-code Solution &
    Not detected \\[0.4em]

\end{tabular}

\caption{Results after 7 days against various vendors.}
\label{tab:results}
\end{table}

\noindent{ \bf{Discussion.}} Our empirical study identified 7 cloning mechanisms and revealed that none them could be detected by the Virustotal vendors within 7 days. This section will discuss our findings in depth, in addition to limitations and challenges. 

\noindent{ \bf{Why is detection accuracy of cloning attacks 0\%?}}
We identify two main reasons. (1) A common tactic employed by most of the cloning attacks is the lack of suspicious text or typography mistakes on the page which may have been a significant factor in avoiding detection. Proposed ML algorithms for phishing detection usually use natural language processing to scan the webpage text that prompts users to input sensitive information like ``password'' and ``credit card'' \cite{xiang2011cantina+}\cite{ding2019keyword}\cite{l2010latent}. 
(2) The URLs we hosted on also might play a factor. Malicious URLs will have words like "login", "submit" or "secure", or use multiple TLDs (e.g. paypal.assistance-form.com), in an attempt to confuse users. 
Our URLs were of the form "https://xxxx-xxxx-yyyy.netlify.app" (where xxxx is a random English word, and yyyy is a string of random letters and numbers), which may seem suspicious due to randomness, and multiple "-" characters. However, unlike other phishing URLs we observed, our URLs had no confusing TLDs and no misleading keywords. The use of HTTPS (as mandated by our ethical agreement) may also have been significant, as phishing URLs can use HTTPS to avoid detection~\cite{patil2019methodical}.



\noindent{\bf{Limitations and Challenges.}}
Our empirical study has the following limitations and challenges. (1) The limitations of VirusTotal in our experiment might affect the results. The results may be inconsistent between the VirusTotal API, and the results of the each individual security vendor's scan. We also scanned VirusTotal for 7 days to ensure the phishing labels on URLs update, which might need 4 days as stated by previous studies \cite{peng2019opening}. (2) Some security vendors might rely on manual user reports, rather than any automated analysis. Our ethical conditions prevented us from targeting real users, eliminating this method of detection. (3) We only targeted PayPal, as we did not have clearance from other top brands. 
\section{Recommendations}\label{sec:rec}
Based on our study, we propose two ways for victim brands to harden their websites against cloning. We also recommend two methods with which vendors may improve the criteria for their ML-based static analysis detectors.

\vspace{0.5em}

\noindent{\ballnumber{1} \bf{Adopting Dynamic Features.}} Cloning attacks typically rely on taking the HTML, CSS, and other static assets (e.g. images) of a legitimate site. Some techniques we identified failed to clone aspects of webpages that are dynamic and rely on server interaction. Thus, developers can mitigate cloning of their sites by ensuring that the input fields for sensitive information only appear in the DOM once users have had some form of interaction with the server. This will increase the burden for attackers attempting a viable cloning attack.

\noindent{\ballnumber{2} \bf{Including External Authentication.}} We found that external authentication services were a barrier to cloning. One example is third-party SSO, where a relying party (e.g. PayPal) uses an external identity provider (IDP) to authenticate and use their services \cite{vapen2016look}. The relying party will use any accounts and information managed by the IDP to provide services to the user. This form of authentication reduces the chances of efficacy of cloning attacks, since malicious sites cannot integrate with such trusted services.

\noindent{\ballnumber{3} \bf{Identifying High Similarity to wider Benign Sites.}} One way to detect if a page is cloned is to compare it to a legitimate site. Similarity-based detection mechanisms have been researched before, with some studies using CSS files \cite{mao2017phishing} or brand logos\cite{lin2021phishpedia}, however these can be circumvented if the attackers don't use particular assets. Others have taken advantage of widely available and exploited phishing kits, and used the similarity of suspicious web-pages to other previously identified phishing sites \cite{xiang2011cantina+}. We propose that anti-phishing scanners store copies of benign sites that are commonly targeted for phishing, and compare the DOM similarity of potential phishing pages with the stored ones.

\noindent{\ballnumber{4} \bf{Detecting Excessive JavaScript.}}
One of the discovered cloning mechanisms relied heavily on JavaScript to dynamically generate HTML. By using regular expressions to find HTML tags in our files, we estimated that samples using this cloning technique ranged from having 10.04\% of all the characters in their source code being HTML tags, to 30.7\%. To better detect this attack, there is a need for ML-based defences that can establish a threshold baseline for  web-pages with a high JavaScript/HTML ratio in their source code. 





\section{Conclusion}
Web phishing is one of the most prevalent cybersecurity threats today. Cloning benign sites is a new strategy in this threat landscape. Not only does it accurately imitate the target site with ease, but it can also evade automatic anti-phishing detection systems. In our study, we collected data on phishing pages, discovered the prevalence of cloning, and identified seven different mechanisms being used by attackers. We then created our own cloned sites, submitting them to VirusTotal, Sucuri and Google Safe Browsing, to be scanned. None of the vendors we checked labelled our sites as malicious, showing that this kind of attack can successfully evade current anti-phishing systems and remain undetected. We offer four recommendations for both legitimate web developers, and the anti-phishing research community, to help mitigate this alarming threat now and into the future.

\section*{Acknowledgment}

The work has been supported by the Cyber Security Research Centre Limited whose activities are partially funded by the Australian Government’s Cooperative Research Centres Programme. This work was also supported in part by the ITRC support program (IITP-2019-2015-0-00403). The authors would like to thank all the anonymous reviewers for their valuable feedback.

\bibliographystyle{unsrt}
\bibliography{references}

\end{document}